\documentclass[preprint,12pt]{elsarticle}

\usepackage{amssymb}

\begin{document}
\begin{frontmatter}


\title{{\bf Quantum public-key cryptosystems based on induced trapdoor one-way transformations}}
\author[label1]{Li Yang}\ead{yangli@gucas.ac.cn}
\author[label1]{Min Liang}
\author[label1]{Bao Li}
\author[label1]{Lei Hu}
\author[label2]{Deng-Guo Feng}
\address[label1]{State Key Laboratory of Information Security, Graduate University of Chinese Academy of Sciences, Beijing 100049, China}
\address[label2]{State Key Laboratory of Information Security, Institute of Software, Chinese Academy of Sciences, Beijing 100190, China}

\begin{abstract}
A concept named induced trapdoor one-way quantum transformation (OWQT) has been introduced, and a theoretical framework of public-key encryption (PKE) of quantum message is presented based on it. Then several kinds of quantum public-key encryption (QPKE) protocols, such as quantum version PKE
of RSA, ElGamal, Goldwasser-Micali, elliptic curve, McEliece, Niederreiter and Okamoto-Tanaka-Uchiyama,
are given within this framework. Though all of these protocols are only computationally secure,
the last three are probably secure in post-quantum era. Besides, theoretical frameworks for public-key authentication and signature of quantum message are also given based on the induced trapdoor OWQT.
As examples, a public-key authentication protocol of quantum message based on SN-S authentication scheme and two quantum digital signature protocols based on RSA and McEliece algorithms respectively are presented.
\end{abstract}


\end{frontmatter}
\journal{}

\section{Introduction}
Most public-key cryptosystems currently used are based on the hardness of problems such as integer factoring and discrete logarithms. Since these problems would not maintain their hardness in post-quantum era~\cite{Shor94}, people have to consider cryptosystems based on other hard problems. It is believed that there does not exist efficient quantum algorithm to solve NP-complete problems~\cite{Bennett97}, therefore, cryptosystems based on NP-complete problems are regarded as good choices against quantum attacks.

Okamoto et al.~\cite{Okamoto00} constructed the first quantum public-key cryptosystem (QPKC) based on subset-sum problem. Their key-generation algorithms include a quantum algorithm, though the private-key, public-key, plaintext and ciphertext are all classical. Gottesman and Chuang~\cite{Gottesman01} constructed a quantum digital signature, whose pubic key is quantum, but private-key and message are classical. In~\cite{Kawachi05}, a QPKC is constructed based on a hard problem so called $QSCD_{ff}$, which has been proved to be one with bounded information theoretic security. By using single-qubit rotations, Nikolopoulos~\cite{Nikolopoulos08} proposed a QPKC with classical private-key and quantum public-key. Based on quantum encryption, Gao et al.~\cite{Gao09} presented a QPKC with symmetric keys, here two qubits from a Bell state serve as the public-key and the private-key respectively. Pan and Yang~\cite{Pan10} constructed a quantum public-key encryption (QPKE) scheme with information theoretic security. These QPKCs are all classical bits oriented.

Yang~\cite{Yang05} proposed a QPKE scheme for quantum message encryption, which is a variation of McEliece public-key cryptosystem~\cite{McEliece78}. In~\cite{Barnum02}, quantum message authentication schemes were discussed. Based on classical SN-S authentication code, a public-key authentication scheme of quantum message was also constructed~\cite{Yang03}.

This paper focuses on the public-key encryption (PKE), authentication and signature of quantum message. A concept named induced trapdoor one-way quantum transformation (OWQT) is introduced, and a computationally secure theoretical framework is presented based on it. QPKE protocols such as quantum version of RSA, ElGamal, Goldwasser-Micali, elliptic curve, McEliece, Niederreiter and Okamoto-Tanaka-Uchiyama PKE are given. Besides, theoretical frameworks for public-key authentication and signature of quantum message are also proposed.

\section{Induced trapdoor one-way quantum transformation}
Quantum transformation $U_f$ computing a function $f:\{0,1\}^n\rightarrow\{0,1\}^m$ is defined as
\begin{equation}U_f\left(|x\rangle|y\rangle\right)=|x\rangle|y\oplus f(x)\rangle,\end{equation}
where $\oplus$ denotes bitwise addition in $\mathcal{F}_2$.

It is worth to mention that the quantum transformation $U_{f^{-1}}$ computing $f^{-1}$ does not equal to $U_f^{-1}$ computing the inverse of $U_f$.

Given function $f(m,r)$, a unitary transformation computing $f$ is defined as
\begin{equation}U_f\left(|r\rangle|m\rangle|0\rangle\right)=|r\rangle|m\rangle|f(m,r)\rangle.\end{equation}
Another unitary transformation $U(f,g)$ computing $m$ from values of $f(m,r)$, $g(m,r)$ and $r$ is defined as
\begin{equation}U(f,g)\left(|r\rangle|0\rangle|g(m,r)\rangle|f(m,r)\rangle\right)=|r\rangle|m\rangle|g(m,r)\rangle|f(m,r)\rangle.\end{equation}

\begin{figure}[htp!]
\begin{center}
\includegraphics[width=10cm]{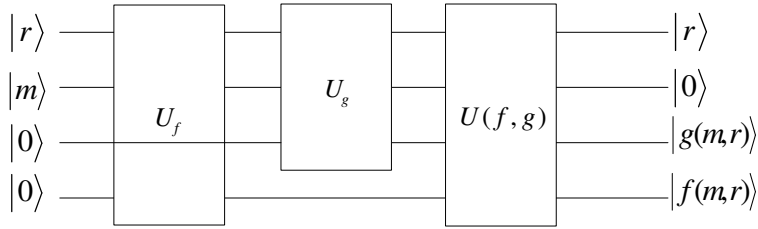}
\end{center}
\caption{\label{fig4} The quantum circuit implementation of $U_{fg}(r)$ via $U_f$,$U_g$,$U(f,g)$. The quantum circuits $U_f$ and $U_g$ compute the functions $f(m,r)$ and $g(m,r)$ respectively. The quantum circuit $U(f,g)$ computes $m$ from $r$, $g(m,r)$ and $f(m,r)$.}
\end{figure}

Unitary transformation implemented via quantum circuits of $U_f$, $U_g$ and $U(f,g)$ is shown in Figure~\ref{fig4}.

It can be seen that the quantum circuit in Figure~\ref{fig4} implements a unitary transformation defined as \begin{equation}U_{fg}(r)\left(|m\rangle|0\rangle|0\rangle\right)=|0\rangle|g(m,r)\rangle|f(m,r)\rangle,\end{equation}
where $g(m,r)\neq g(m',r)$ and $f(m,r)\neq f(m',r)$ if $m\neq m'$. To the receiver and adversaries, this transformation can be regarded as a trace-preserving quantum operation.

{\bf Definition 1:} Given a classical trapdoor one-way function $f(m,r)$ with a random parameter $r$, and a classical function $g(m,r)$, the quantum transformation $U_{fg}(r):|m\rangle\rightarrow|g(m,r)\rangle|f(m,r)\rangle$ is an induced trapdoor one-way quantum transformation if it satisfy
\begin{description}
    \item[\it{1. Easy to operate.}] A sufficient condition is: both $f(m,r)$ and $g(m,r)$ can be computed efficiently; Given $r$, one can efficiently get $m$ from $f(m,r)$ or $g(m,r)$.
    \item[\it{2. Hard to invert.}] A sufficient condition is: from the values of $f(m,r)$ and $g(m,r)$, one cannot efficiently get both $m$ and $r$.
    \item[\it{3. Easy to invert with the trapdoor $s$.}] A sufficient condition is: with the trapdoor $s$, one can efficiently get $m$ from $f(m,r)$ and $g(m,r)$, and efficiently get $r$ from $m$, $f(m,r)$ and $g(m,r)$.
\end{description}

{\bf Remark 1:} In "1", it is required that $m$ can be efficiently obtained from $r$, $f(m,r)$ and $g(m,r)$. This condition is necessary for the implementation of the quantum transformation $U_{fg}(r)$, see Figure~\ref{fig4}. The property 2 means that the adversary without $r$ cannot get $U_{fg}^\dagger(r)$. In "3",  for the case that $r$ cannot be obtained even with the aid of trapdoor $s$, we have to require that 1) $g(m,r)=\tilde{g}(r)$ or $g(m,r)=\tilde{g}(m)$; 2) $f(m,r)$ can be efficiently evaluated from $s$, $m$ and $g(m,r)$.

\section{Public-key cryptosystems of quantum message}
\subsection{Public-key encryption}
Consider encrypting a quantum message $\sum_m\alpha_m|m\rangle$ with induced trapdoor OWQT $U_{fg}(r)$. The algorithm is as follows:
\begin{eqnarray}
&&|r\rangle\sum_m\alpha_m|m\rangle|0\rangle|0\rangle \nonumber\\
& \stackrel{1}{\rightarrow} & |r\rangle\sum_m\alpha_m|m\rangle|g(m,r)\rangle|f(m,r)\rangle \nonumber\\
& \stackrel{2}{\rightarrow} & |r\rangle|0\rangle\sum_m\alpha_m|g(m,r)\rangle|f(m,r)\rangle,
\end{eqnarray}
which completes the encryption transformation
\begin{equation}
U_{fg}(r)\left(\sum_m\alpha_m|m\rangle|0\rangle|0\rangle\right)=|0\rangle\sum_m\alpha_m|g(m,r)\rangle|f(m,r)\rangle.
\end{equation}
According to the definition of induced trapdoor OWQT, the quantum transformation $U_{fg}(r)$ is an efficient encryption transformation. It can be seen that, given the value of $r$, the inverse transformation of $U_{fg}(r)$ can also be operated efficiently.

Because Bob do not know the value of $r$, the quantum cipher state to him is a mixed state with density matrix
\begin{equation}
\sum_rp_r(\sum_m\alpha_m|g(m,r)\rangle|f(m,r)\rangle)(\sum_m\alpha_m^*\langle g(m,r)|\langle f(m,r)|).
\end{equation}

Given the trapdoor $s$ of $f(m,r)$, the decryption transformation on quantum cipher state $\sum_m\alpha_m|g(m,r)\rangle|f(m,r)\rangle$ proceeds as follows (without loss of generality, we restrict our attention to a pure state in the decryption procedure).

For the case that $r$ cannot be obtained, we require $g(m,r)$ depending only on $m$ or $r$ (according to the definition of induced trapdoor OWQT, $g(m,r)=\tilde{g}(r)$ or $g(m,r)=\tilde{g}(m)$), and the decryption is as follows:

\begin{eqnarray}\label{equ0}
&&|s\rangle|0\rangle\sum_m\alpha_m|\tilde{g}(r)\rangle|f(m,r)\rangle \nonumber\\
&\stackrel{1}{\rightarrow} & |s\rangle\sum_m\alpha_m|m\rangle|\tilde{g}(r)\rangle|f(m,r)\rangle \nonumber\\
&\stackrel{2}{\rightarrow} & |s\rangle\sum_m\alpha_m|m\rangle|\tilde{g}(r)\rangle|0\rangle. \nonumber\\
\end{eqnarray}
or
\begin{eqnarray}\label{equ1}
&&|s\rangle|0\rangle\sum_m\alpha_m|\tilde{g}(m)\rangle|f(m,r)\rangle \nonumber\\
&\stackrel{1}{\rightarrow} & |s\rangle\sum_m\alpha_m|m\rangle|\tilde{g}(m)\rangle|f(m,r)\rangle \nonumber\\
&\stackrel{2}{\rightarrow} & |s\rangle\sum_m\alpha_m|m\rangle|\tilde{g}(m)\rangle|0\rangle \nonumber\\
&\stackrel{2}{\rightarrow} & |s\rangle\sum_m\alpha_m|m\rangle|0\rangle|0\rangle.
\end{eqnarray}
Suppose $m$ can be efficiently get from the value of $f(m,r)$ and $g(m,r)$ with the trapdoor $s$ (see the sufficient condition of "3" in the definition of $U_{fg}(r)$), the first step can be carried out efficiently. If $f(m,r)$ can be efficiently computed from $s$, $m$ and $\tilde{g}(r)$, the second step can also be carried out efficiently (see "1","3" and Remark 1).

For the case that $r$ can be obtained with the trapdoor $s$, the decryption is as follows:
\begin{eqnarray}\label{equ2}
&&|s\rangle|0\rangle|0\rangle\sum_m\alpha_m|g(m,r)\rangle|f(m,r)\rangle \nonumber\\
&\stackrel{1}{\rightarrow} & |s\rangle|r\rangle\sum_m\alpha_m|m\rangle|g(m,r)\rangle|f(m,r)\rangle \nonumber\\
&\stackrel{2}{\rightarrow} & |s\rangle|r\rangle\sum_m\alpha_m|m\rangle|0\rangle|0\rangle.
\end{eqnarray}
In the above two steps, the first step can be carried out efficiently according to the property "3", and the the quantum transformations $U_f$ and $U_g$ are efficiently performed in the second step. Then the quantum message $\sum_m\alpha_m|m\rangle$ can be obtained after polynomial time quantum computation. Denote the decryption transformation as $D_{1s}(f,g)$ and $D_{2s}(f,g)$ for case 1 and case 2, respectively. The decryption transformations are as follows:
\begin{eqnarray}
&D_{1s}(f,g)\left(|0\rangle\sum_m\alpha_m|g(r)\rangle|f(m,r)\rangle\right)=\sum_m\alpha_m|m\rangle|g(r)\rangle|0\rangle, \\
or &D_{1s}(f,g)\left(|0\rangle\sum_m\alpha_m|g(m)\rangle|f(m,r)\rangle\right)=\sum_m\alpha_m|m\rangle|0\rangle|0\rangle,
\end{eqnarray}
\begin{equation}
D_{2s}(f,g)\left(|0\rangle|0\rangle\sum_m\alpha_m|g(m,r)\rangle|f(m,r)\rangle\right)=|r\rangle\sum_m\alpha_m|m\rangle|0\rangle|0\rangle.
\end{equation}
Then we arrive at the following protocol:

$f(m,r)$ is a trapdoor one-way function, and Bob posses its trapdoor $s$. $f(m,r)$ and $g(m,r)$ are public.
\begin{description}
    \item[{\it Ecryption}] To encrypt a quantum message $\sum_m\alpha_m|m\rangle$, Alice selects randomly a number $r$, then carries out the encryption transformation $U_{fg}(r)$, and obtained the cipher state $\sum_m\alpha_m|g(m,r)\rangle|f(m,r)\rangle$. Then she sends the cipher state to Bob (Notice that classical plaintext communication is allowed here).
    \item[{\it Decryption}] Bob performs the decryption transformation $D_{1s}(f,g)$ or $D_{2s}(f,g)$ to the cipher state, and get the quantum message $\sum_m\alpha_m|m\rangle$.
\end{description}

\subsection{\label{seccount3}Authentication}
In a classical authentication scheme, the authentication rule is $h(m)=(m,a(m))$, here $a(m)$ is the authentication code of message $m$. An authentication scheme for quantum message can be described as follows:

(1) Alice encodes a $k$-qubit message $\sum_m\alpha_m|m\rangle$ as follows:
\begin{eqnarray}
&&\sum_m\alpha_m|m\rangle|0\rangle \nonumber\\
&\rightarrow& \sum_m\alpha_m|m\rangle|h(m)\rangle=\sum_m\alpha_m|m\rangle|m,a(m)\rangle \nonumber\\
&\rightarrow& |0\rangle\sum_m\alpha_m|m,a(m)\rangle.
\end{eqnarray}

(2) Alice encrypts the quantum state $\sum_m\alpha_m|m,a(m)\rangle$ via PKE of quantum message.

(3) Bob decrypts the received quantum state and obtains the plaintext $\sum_m\alpha_m|m,a(m)\rangle$.

(4) Bob carries out the following transformation to the quantum state $\sum_m\alpha_m|m,a(m)\rangle$.
\begin{eqnarray}
&&\sum_m\alpha_m|m,a(m)\rangle|0\rangle \nonumber\\
&\rightarrow& \sum_m\alpha_m|m,a(m)\rangle|m\rangle \nonumber\\
&\rightarrow& \sum_m\alpha_m|0,a(m)\rangle|m\rangle \nonumber\\
&\rightarrow& \sum_m\alpha_m|0,a(m)\oplus a(m)\rangle|m\rangle=|0\rangle\sum_m\alpha_m|m\rangle.
\end{eqnarray}

(5) Bob measures the first register to check whether it is in the state $|0\rangle$, then he gets the message coming from Alice in the second register with authentication.

In this kind of authentication scheme of quantum message, the authentication rule $h(m)$ is public and the scheme is a public-key data integrity scheme.

{\bf Remark 2:} If we require the scheme to be one against substitution, it should be modified slightly as follows: Suppose Alice's identity information $S$ cannot be forged. A quantum register named identity register is initiated with quantum state $|S\rangle$. In step (1), Alice firstly carries out an Hadamard transformation $H^{\otimes l}$ on the quantum state $|S\rangle$, then encodes the quantum state $H^{\otimes l}(|S\rangle)\sum_m\alpha_m|m\rangle$. In step (5), Bob finally obtains the state $H^{\otimes l}(|S\rangle)\sum_m\alpha_m|m\rangle$. After step (5), he carries out Hadamard transformation $H^{\otimes l}$ on state $H^{\otimes l}(|S\rangle)$ and gets $|S\rangle$, then measures it to identify the sender. Since the identity information $S$ cannot be forged, the attackers cannot substitute the message successfully.

\subsection{\label{seccount6}Digital signature}
Suppose $f:\{0,1\}^{k+n}\rightarrow\{0,1\}^{k'+n'}$ is a trapdoor one-way function, Alice has its trapdoor $s$. Alice signs a quantum message $\sum_m\alpha_m|m\rangle$ to Bob as follows:

(1) Bob randomly generates a number $r_B\in\{0,1\}^{k'}$, and sends it to Alice.

(2) Alice randomly generates a number $r_A\in\{0,1\}^{n'}$, and computes
\begin{equation}f^{-1}(r_B,r_A)=(r,r'),\end{equation}where $r\in\{0,1\}^k$ and $r'\in\{0,1\}^n$.
Then Alice signs the quantum message $\sum_m\alpha_m|m\rangle$
\begin{equation}\sum_m\alpha_m|m\rangle\rightarrow\sum_m\alpha_m|m\rangle|f(m,r)\rangle,\end{equation}
and sends the quantum state $\sum_m\alpha_m|m\rangle|f(m,r)\rangle$ to Bob.

(3) Bob tells Alice that he has received the quantum state.

(4) Alice announces $r$ and $r'$.

(5) Bob computes $f(r,r')$ and checks whether the first $k'$ bits are $r_B$.
Then he performs the transformation
\begin{equation}\sum_m\alpha_m|m\rangle|f(m,r)\rangle\rightarrow\sum_m\alpha_m|m\rangle|0\rangle,\end{equation}
and measures the second quantum register. He accepts the signature if and only if the second register is in state $|0\rangle$.

{\bf Remark 3:} (1) These protocols are interactive digital signature protocols of quantum message. (2) They are undeniable signature protocols and Alice's collaboration is needed during the verification. (3) Multiple verification is possible through copying $|f(m,r)\rangle$ to other registers. But after the quantum message $\sum_m\alpha_m|m\rangle$ is extracted, it is impossible to verify any more. So these signatures are signed on the envelop and this kind of signature should be termed as "quantum sealing wax".

\section{Concrete protocols}
A quantum message is a sequence of pure states. Without loss of generality, we restrict our
attention to the encryption and decryption of a pure state.
\subsection{\label{seccount1}Encryption protocols without post-quantum security}
\subsubsection{\label{seccount4}Quantum RSA PKE}
In RSA PKE~\cite{Rivest78}, $p$ and $q$ are two large primes, $N=pq$, $\phi(N)=(p-1)(q-1)$, $e$ satisfies $(e,\phi(N))=1$, and $s=e^{-1}\textrm{mod}(\phi(N))$.
According to the theoretical framework established in the previous section, we construct a PKE of quantum message which is a quantum version of RSA.
Let $g(m,r)=m\oplus r$, $f(m,r)=m^e \textrm{mod} N$, $s$ is the trapdoor of $f(m,r)$.

{\bf Encryption}\\

Alice selects a value of $r$, then does the following encryption transformation
\begin{eqnarray}
&&|r\rangle\sum_m\alpha_m|m\rangle|0\rangle \nonumber\\
& \rightarrow & |r\rangle\sum_m\alpha_m|m\rangle|m^e \textrm{mod} N\rangle \nonumber\\
& \rightarrow & |r\rangle\sum_m\alpha_m|m\oplus r\rangle|m^e \textrm{mod} N\rangle.
\end{eqnarray}
After that, she sends to Bob the cipher state $\sum_m\alpha_m|m\oplus r\rangle|m^e \textrm{mod} N\rangle$.

\vspace{3mm}
{\bf Decryption}\\

After receiving the cipher state, Bob does the decryption transformation using the private-key $s$,
\begin{eqnarray}
&&|s\rangle\sum_m\alpha_m|m\oplus r\rangle|m^e \textrm{mod} N\rangle|0\rangle \nonumber\\
& \rightarrow & |s\rangle\sum_m\alpha_m|m\oplus r\rangle|m^e \textrm{mod} N\rangle|(m^e)^s \textrm{mod} N\rangle \nonumber\\
&&=|s\rangle\sum_m\alpha_m|m\oplus r\rangle|m^e \textrm{mod} N\rangle|m\rangle \nonumber\\
& \rightarrow & |s\rangle|r\rangle\sum_m\alpha_m|m^e \textrm{mod} N\rangle|m\rangle \nonumber\\
& \rightarrow & |s\rangle|r\rangle|0\rangle\sum_m\alpha_m|m\rangle.
\end{eqnarray}
Finally, Bob obtains the quantum message $\sum_m\alpha_m|m\rangle$.

\subsubsection{Quantum ElGamal PKE}
In the ElGamal PKE \cite{ElGamal85}, $s$ is private, $p,\alpha,\beta$ are public, here $\beta=\alpha^s$. Let $g(m,r)=\alpha^r\textrm{mod}p$ and $f(m,r)=m\beta^r\textrm{mod}p$. The quantum ElGamal PKE is as follows:

{\bf Encryption}\\

Alice randomly selects a number $r$ and performs the following transformations to encrypt a quantum message $\sum_m\alpha_m|m\rangle$:
\begin{eqnarray}
&&|r\rangle\sum_m\alpha_m|m\rangle|0\rangle \nonumber\\
&\rightarrow& |r\rangle\sum_m\alpha_m|m\rangle|m\beta^r \textrm{mod} p\rangle.
\end{eqnarray}
Then Alice sends $\alpha^r\textrm{mod}p$ and the cipher state $\sum_m\alpha_m|m\rangle|m\beta^r \textrm{mod} p\rangle$ to Bob.

\vspace{3mm}
{\bf Decryption}\\

After receiving the cipher state and $\alpha^r\textrm{mod}p$, Bob decrypts it using the private-key $s$. The procedure is as follows:
\begin{eqnarray}
&&|s\rangle|\alpha^r\textrm{mod}p\rangle\sum_m\alpha_m|m\rangle|m\beta^r \textrm{mod} p\rangle \nonumber\\
&\rightarrow& |s\rangle|\alpha^r\textrm{mod}p\rangle\sum_m\alpha_m|m\rangle|m\beta^r\oplus m(\alpha^r)^s \textrm{mod} p\rangle \nonumber\\
&&=|s\rangle|\alpha^r\textrm{mod}p\rangle\sum_m\alpha_m|m\rangle|0\rangle.
\end{eqnarray}
Then Bob obtains the quantum message $\sum_m\alpha_m|m\rangle$.

\subsubsection{Quantum Goldwasser-Micali PKE}
In Goldwasser-Micali PKE \cite{Goldwasser84}, $p$ and $q$ are two primes, $N=pq$, $t\in \mathcal{Z}_N^1$
is a quadratic nonresidue modulo $N$. $N,t$ are public and $p,q$ are private. $Q_N(x)=1$ if $x$
is a quadratic residue modulo $N$, otherwise $Q_N(x)=0$. To encrypt a binary string $m=m_1m_2\cdots m_k$,
Alice selects randomly $r_1,r_2,\ldots,r_k$, then computes $c_i=t^{m_i}r_i^2 \textrm{mod} N$ for $i=1,2,\ldots,k$.
The numbers $(c_1,c_2,\ldots,c_k)$ are sent to Bob as the cipher. As Bob knows the factors of $N$,
he can know whether $c_i$ is a quadratic residue modulo $N$. Let $m_i=Q_N(c_i)$, he obtains the plaintext
$m=m_1\cdots m_k$.

Let $g(m,r_1,\cdots,r_k)=\left( m\oplus r_1, (r_1m \textrm{mod}2^k) \oplus r_2,
\ldots, (r_{k-1}m\textrm{mod}2^k) \oplus r_k \right)$ and $f(m,r_1,\cdots,r_k)=( c_1,\cdots,c_k )$,
here $c_i=t^{m_i}r_i^2 \textrm{mod} N$ and $m_i$ is the $i$th bit of its binary string.
The quantum Goldwasser-Micali PKE is as follows:

\vspace{3mm}
{\bf Encryption}\\

Alice encrypts the quantum message $\sum_m\alpha_m|m\rangle$ via computing
\begin{eqnarray}
&&|r_1\cdots r_k\rangle\sum_m\alpha_m|m\rangle|0\rangle|0\rangle \nonumber\\
&\rightarrow & |r_1\cdots r_k\rangle\sum_m\alpha_m|m\rangle|0\rangle|c_1\cdots c_k\rangle \nonumber\\
&\rightarrow & |r_1\cdots r_k\rangle\sum_m\alpha_m|m\rangle|m\oplus r_1, \nonumber\\
&&(r_1m\textrm{mod}2^k) \oplus r_2,\ldots,(r_{k-1}m\textrm{mod}2^k)\oplus r_k\rangle|c_1\cdots c_k\rangle \nonumber\\
&\rightarrow & |r_1\cdots r_k\rangle|0\rangle\sum_m\alpha_m|m\oplus r_1,(r_1m\textrm{mod}2^k)\oplus r_2, \nonumber\\
&&\ldots,(r_{k-1}m\textrm{mod}2^k)\oplus r_k\rangle|c_1\cdots c_k\rangle,
\end{eqnarray}
then sends the cipher state $\sum_m\alpha_m|m\oplus r_1, (r_1m \textrm{mod}2^k) \oplus r_2,
\ldots, (r_{k-1}m\textrm{mod}2^k) \oplus r_k\rangle|c_1\cdots c_k\rangle$ to Bob.

\vspace{3mm}
{\bf Decryption}\\

After receiving the cipher state $\sum_m\alpha_m|m\oplus r_1,(r_1m\textrm{mod}2^k)\oplus r_2,\ldots,
(r_{k-1}m\textrm{mod}2^k) \oplus r_k\rangle|c_1\cdots c_k\rangle$, Bob computes
\begin{eqnarray}
&&|p,q\rangle\sum_m\alpha_m|m\oplus r_1,(r_1m\textrm{mod}2^k)\oplus r_2,\nonumber\\
&&\ldots,(r_{k-1}m\textrm{mod}2^k)\oplus r_k\rangle|c_1\cdots c_k\rangle|0\rangle \nonumber\\
&\rightarrow&|p,q\rangle\sum_m\alpha_m|m\oplus r_1,(r_1m\textrm{mod}2^k)\oplus r_2,\nonumber\\
&&\ldots,(r_{k-1}m\textrm{mod}2^k)\oplus r_k\rangle|c_1\cdots c_k\rangle|m\rangle \nonumber\\
&\rightarrow&|p,q\rangle|r_1,\ldots,r_k\rangle\sum_m\alpha_m|c_1\cdots c_k\rangle|m\rangle \nonumber\\
&\rightarrow&|p,q\rangle|r_1,\ldots,r_k\rangle\sum_m\alpha_m|0\rangle|m\rangle \nonumber\\
&&=|p,q\rangle|r_1,\ldots,r_k\rangle|0\rangle\sum_m\alpha_m|m\rangle.
\end{eqnarray}
Finally, Bob obtains the quantum message $\sum_m\alpha_m|m\rangle$.

\subsubsection{Quantum elliptic curve PKE}
In \cite{Koblitz87}, the classical elliptic curves PKE is proposed. An elliptic curve defined over $\mathcal{Z}_p$ ($p>3$ is prime) is the set of solutions $(x,y)\in\mathcal{Z}_p\times\mathcal{Z}_p$ to the equation
$y^2\equiv x^3+ax+b (\textrm{mod}p)$, here $a,b\in\mathcal{Z}_p$ satisfy $4a^3+27b^2\neq 0 (\textrm{mod}p)$. The points on the elliptic curve form a group
with identity element the point at infinity. Given a point $P$ does not equal to identity element, and chosen $Q$ being $sP$, $s$ is the private-key and $Q$ is the public-key.

Let $g(m,r)=rP$ and $f(m,r)=m\oplus x_2$, here $x_2$ satisfies $(x_2,y_2)=rQ$. The quantum elliptic curve PKE is as follows.

\vspace{3mm}
{\bf Encryption}\\

Alice randomly selects a number $r$, and computes $rQ=(x_2,y_2)$. Given any quantum message $\sum_m\alpha_m|m\rangle$, she carries out encryption with $r$ as follows:
\begin{eqnarray}
&&|r\rangle|0\rangle|0\rangle\sum_m\alpha_m|m\rangle \nonumber\\
&\rightarrow&|r\rangle|x_2,y_2\rangle\sum_m\alpha_m|rP\rangle|m\rangle \nonumber\\
&\rightarrow&|r\rangle|x_2,y_2\rangle|rP\rangle\sum_m\alpha_m|m\oplus x_2\rangle,
\end{eqnarray}
then sends the quantum state $|rP\rangle\sum_m\alpha_m|m\oplus x_2\rangle$.

\vspace{3mm}
{\bf Decryption}\\

Bob receives the cipher state $|rP\rangle\sum_m\alpha_m|m\oplus x_2\rangle$, then uses $s$ to decrypt it:
\begin{eqnarray}
&&|s\rangle|rP\rangle\sum_m\alpha_m|m\oplus x_2\rangle \nonumber\\
&\rightarrow& |s\rangle|x_2,y_2\rangle\sum_m\alpha_m|m\oplus x_2\rangle \nonumber\\
&\rightarrow& |s\rangle|x_2,y_2\rangle\sum_m\alpha_m|m\rangle.
\end{eqnarray}
Finally, Bob obtains the quantum message $\sum_m\alpha_m|m\rangle$. Notice that in the cipher state, $|rP\rangle$ can be replaced with classical message $(x_1,y_1)$.

\subsection{\label{seccount2}Encryption protocols with post-quantum security}
\subsubsection{\label{seccount5}Quantum McEliece PKE~\cite{Yang05}}
Consider McEliece PKE protocol~\cite{McEliece78}. Suppose $G$ is a $k\times n$ generator matrix of a Goppa code,
$G'=SGP$, here $S$ is a $k\times k$ invertible matrix and $P$ is an $n\times n$ permutation matrix.
We choose $G'$ as the public-key and $(S,G, P)$ as the private-key. Let $H$ is the check matrix of Goppa code
satisfying $GH^T=0$. Suppose $g(m,r)=0$ and $f(m,r)=mG'\oplus r$. The quantum McEliece PKE scheme is as follows:

\vspace{3mm}
{\bf Encryption}\\

Alice selects a random number $r$, and uses Bob's public-key $G'$ with $r$ to encrypt a $k$-qubit state $\sum_m\alpha_m|m\rangle$ as follows:
\begin{eqnarray}
&&|r\rangle\sum_m\alpha_m|m\rangle|0\rangle \nonumber\\
&\rightarrow& |r\rangle\sum_m\alpha_m|m\rangle|mG'\rangle \nonumber\\
&\rightarrow& |r\rangle\sum_m\alpha_m|m\oplus mG'G'^{-1}\rangle|mG'\rangle=|r\rangle|0\rangle\sum_m\alpha_m|mG'\rangle \nonumber\\
&\rightarrow& |r\rangle|0\rangle\sum_m\alpha_m|mG'\oplus r\rangle,
\end{eqnarray}
where the matrix $G'^{-1}$ is a generalized inverse matrix of $G'$. Because $G'$ is a full row
rank matrix, there exists $G'^{-1}$ that satisfies $G'G'^{-1}=I_k$. This is the condition that one can get $\sum_m\alpha_m|mG'\rangle$ from $\sum_m\alpha_m|m\rangle$. Alice sends the cipher state $\sum_m\alpha_m|mG'\oplus r\rangle$ to Bob.

\vspace{3mm}
{\bf Decryption}\\

Bob uses his private-key $s=(S,G,P)$ to decrypt the state coming from Alice,
\begin{eqnarray}
&&|s\rangle\sum_m\alpha_m|mG'\oplus r\rangle|0\rangle|0\rangle \nonumber\\
&\rightarrow&|s\rangle\sum_m\alpha_m|mG'\oplus r\rangle|(mG'\oplus r)P^{-1}\rangle|0\rangle \nonumber\\
&\rightarrow&|s\rangle\sum_m\alpha_m|0\rangle|(mG'\oplus r)P^{-1}\rangle|0\rangle=|s\rangle|0\rangle\sum_m\alpha_m|mSG\oplus rP^{-1}\rangle|0\rangle \nonumber\\
&\rightarrow&|s\rangle|0\rangle\sum_m\alpha_m|mSG\oplus rP^{-1}\rangle|(mSG\oplus rP^{-1})H^T\rangle \nonumber\\
&&=|s\rangle|0\rangle\sum_m\alpha_m|mSG\oplus rP^{-1}\rangle|rP^{-1}H^T\rangle,
\end{eqnarray}
then measures the second register to get $rP^{-1}H^T$, and find $rP^{-1}$ via the fast decoding
algorithm of the Goppa code generated by $G$. Bob carries out the following transformation on the quantum state $\sum_m\alpha_m|mSG\oplus rP^{-1}\rangle$ according to the value of $rP^{-1}$,
\begin{equation}
|rP^{-1}\rangle\sum_m\alpha_m|mSG\oplus rP^{-1}\rangle\rightarrow|rP^{-1}\rangle\sum_m\alpha_m|mSG\rangle.
\end{equation}
Then he computes
\begin{eqnarray}
&&|s\rangle\sum_m\alpha_m|mSG\rangle|0\rangle|0\rangle \nonumber\\
&\rightarrow&|s\rangle\sum_m\alpha_m|mSG\rangle|mSGG^{-1}\rangle|0\rangle=|s\rangle\sum_m\alpha_m|mSG\rangle|mS\rangle|0\rangle \nonumber\\
&\rightarrow&|s\rangle\sum_m\alpha_m|0\rangle|mS\rangle|0\rangle \nonumber\\
&\rightarrow&|s\rangle|0\rangle\sum_m\alpha_m|mS\rangle|mSS^{-1}\rangle=|s\rangle|0\rangle\sum_m\alpha_m|mS\rangle|m\rangle \nonumber\\
&\rightarrow&|s\rangle|0\rangle|0\rangle\sum_m\alpha_m|m\rangle.
\end{eqnarray}
Finally, the quantum message $\sum_m\alpha_m|m\rangle$ is obtained.

\subsubsection{Quantum Niederreiter PKE}
In Niederreiter PKE protocol~\cite{Niederreiter86}, $M$ is an invertible matrix,
$H$ is a check matrix of a code with random-error-correcting capability $t$,
and $P$ is a permutation matrix. Let $H'=MHP$. $(M,H,P)$ is the private-key and $H'$ is the public-key.
Let $g(m,r)=m\oplus r$, $f(m,r)=mH'^T$, the quantum Niederreiter PKE is as follows:

\vspace{3mm}
{\bf Encryption}\\

Alice randomly selects an error vector $r$ which satisfies $w(r)=t$, here $w(\cdot)$ represents Hamming weight. She encrypts a quantum message $\sum_m\alpha_m|m\rangle$ using $r$:
\begin{eqnarray}
&&|r\rangle\sum_m\alpha_m|m\rangle|0\rangle \nonumber\\
&\rightarrow& |r\rangle\sum_m\alpha_m|m\rangle|mH'^T\rangle \nonumber\\
&\rightarrow& |r\rangle\sum_m\alpha_m|m\oplus r\rangle|mH'^T\rangle,
\end{eqnarray}
then sends the quantum states $\sum_m\alpha_m|m\oplus r\rangle|mH'^T\rangle$ as cipher state to Bob.

\vspace{3mm}
{\bf Decryption}\\

Bob receives the cipher state and decrypts it as follows: he computes
\begin{eqnarray}
&&\sum_m\alpha_m|m\oplus r\rangle|mH'^T\rangle \nonumber\\
&\rightarrow&\sum_m\alpha_m|m\oplus r\rangle|mH'^T\oplus (m\oplus r)H'^T\rangle \nonumber\\
&&=\sum_m\alpha_m|m\oplus r\rangle|rH'^T\rangle,
\end{eqnarray}
and then uses the private-key $s=(M,H,P)$ to computes $r$ which includes 4 steps 1) measure the second register and obtain $rH'^T$; 2) compute $rH'^T(M^T)^{-1}=r(MHP)^T(M^T)^{-1}=rP^TH^T$;
3) find $rP^T$ via the fast decoding algorithm of the code generated by $H$; 4) compute $(rP^T)(P^T)^{-1}=r$. Finally, he performs the following transformation according to the value of $r$:
\begin{equation}
|r\rangle\sum_m\alpha_m|m\oplus r\rangle \rightarrow |r\rangle\sum_m\alpha_m|m\rangle,
\end{equation}
and obtains the quantum message $\sum_m\alpha_m|m\rangle$.

\subsubsection{Quantum Okamoto-Tanaka-Uchiyama PKE}
In the Okamoto-Tanaka-Uchiyama PKE scheme \cite{Okamoto00}, $(g,d,p,p_1,p_2,\ldots,p_n)$ is private-key.
The public-key $(n,k,b_1,b_2,\ldots,b_n)$ is computed from the private-key with Shor's algorithm for finding
discrete logarithms~\cite{Shor94}. In the encryption procedure, the plaintext $m$ is encoded to a
code $e(m)=e_1e_2\cdots e_n$ of constant weight $k$, the cipher is $c(m)=\sum_{i=1}^ne_ib_i$.
In the decryption procedure, Bob computes $u=g^{(c-kd)\textrm{mod}(p-1)}\textrm{mod}p$,
then chooses $e_i=1$ if $p_i|u$, otherwise $0$. Finally, he computes $m=e_i\sum_{i=1}^nC_{n-i}^{k-\sum_{j=1}^{i-1}e_j}$.

Let $g(m,r)=m\oplus r$ and $f(m,r)=\tilde{f}(m)=\sum_{i=1}^ne_ib_i$, here $e_1\cdots e_n$ is the constant weight code of $m$. We construct a quantum Okamoto-Tanaka-Uchiyama PKE as follows.

\vspace{3mm}
{\bf Encryption}\\

Alice randomly selects a number $r$, then encrypts the quantum message $\sum_m\alpha_m|m\rangle$
using $r$ and the public-key $(n,k,b_1,b_2,\ldots,b_n)$. Suppose $e(m)=e_1e_2\cdots e_n$ is the constant weight encoding
of $m$, and $c(m)=\sum_{i=1}^ne_ib_i$ is the cipher of $m$. Alice computes
\begin{eqnarray}
&&|r\rangle\sum_m\alpha_m|m\rangle|0\rangle|0\rangle \nonumber\\
&\rightarrow& |r\rangle\sum_m\alpha_m|m\rangle|0\rangle|e(m)\rangle \nonumber\\
&\rightarrow& |r\rangle\sum_m\alpha_m|m\rangle|c(m)\rangle|e(m)\rangle \nonumber\\
&\rightarrow& |r\rangle\sum_m\alpha_m|m\rangle|c(m)\rangle|0\rangle \nonumber\\
&\rightarrow& |r\rangle\sum_m\alpha_m|m\oplus r\rangle|c(m)\rangle|0\rangle,
\end{eqnarray}
then obtains the cipher state $\sum_m\alpha_m|m\oplus r\rangle|c(m)\rangle$.

\vspace{3mm}
{\bf Decryption}\\

Bob uses his private-key $s=(g,d,p,p_1,p_2,\ldots,p_n)$ to decrypt the cipher state.
During the decryption process, in order to get $e(m)$ from $c(m)$, Bob computes $u=g^{(c(m)-kd)\textrm{mod}(p-1)}\textrm{mod}p$ firstly,
then check if $p_i|u$ for each $i\in\{1,2,\cdots,n\}$. If $p_i|u$, then set $e_i=1$, otherwise, set $e_i=0$. Based on this
algorithm, he can computes
\begin{eqnarray}
&&|s\rangle\sum_m\alpha_m|m\oplus r\rangle|c(m)\rangle|0\rangle|0\rangle \nonumber\\
&\rightarrow&|s\rangle\sum_m\alpha_m|m\oplus r\rangle|c(m)\rangle|e(m)\rangle|0\rangle \nonumber\\
&\rightarrow&|s\rangle\sum_m\alpha_m|m\oplus r\rangle|c(m)\rangle|e(m)\rangle|m\rangle \nonumber\\
&\rightarrow&|s\rangle|r\rangle\sum_m\alpha_m|c(m)\rangle|e(m)\rangle|m\rangle \nonumber\\
&\rightarrow&|s\rangle|r\rangle|0\rangle\sum_m\alpha_m|e(m)\rangle|m\rangle \nonumber\\
&\rightarrow&|s\rangle|r\rangle|0\rangle|0\rangle\sum_m\alpha_m|m\rangle.
\end{eqnarray}
Finally, he obtains the quantum message $\sum_m\alpha_m|m\rangle$.

\subsection{Remarks of QPKE protocols}
We have proposed seven QPKE protocols, which are all under our theoretical framework.
The four protocols in Sec.\ref{seccount1} are based on factoring problem or discrete
logarithms problem which can be solved efficiently on quantum computer. However,
these protocols can help us to understand the theoretical framework of quantum message oriented PKE.
The three protocols in Sec.\ref{seccount2} are based on the hardness of NP-complete problem and currently
regarded as ones with post-quantum security.

In this section, we give a brief overview of the above seven protocols.

\vspace{3mm}
(1) Quantum RSA PKE\\

$g(m,r)=m\oplus r$, $f(m,r)=m^e \textrm{mod} N$, and the trapdoor is $s=e^{-1}\textrm{mod}(\phi(N))$.

\vspace{3mm}
(2) Quantum ElGamal PKE\\

$g(m,r)=m$, $f(m,r)=m\beta^r\textrm{mod}p$, and the trapdoor $s$ satisfies $\beta=\alpha^s$. In this protocol, classical message $\alpha^r\textrm{mod}p$ must be transmitted.

\vspace{3mm}
(3) Quantum Goldwasser-Micali PKE\\

$g(m,r_1,\cdots,r_k)=\left( m\oplus r_1, (r_1m \textrm{mod}2^k) \oplus r_2,
\ldots, (r_{k-1}m\textrm{mod}2^k) \oplus r_k \right)$ and $f(m,r_1,\cdots,r_k)=( c_1,\cdots,c_k )$, here $c_i=t^{m_i}r_i^2 \textrm{mod} N$ and $m_i$ is the $i$th bit of its binary string. In this protocol, the primes $p,q$ are the trapdoor, which satisfy $pq=N$.

\vspace{3mm}
(4) Quantum elliptic curve PKE\\

$g(m,r)=rP$ and $f(m,r)=m\oplus x_2$, here $x_2$ satisfies $(x_2,*)=rQ$. The trapdoor $s$ satisfies $Q=sP$. In this protocol, $|rP\rangle$ in the cipher state can be replaced with classical message $rP=(x_1,y_1)$.

\vspace{3mm}
(5) Quantum McEliece PKE\\

$g(m,r)=0$ and $f(m,r)=mG'\oplus r$. The trapdoor $s\triangleq(S,G,P)$ satisfies $SGP=G'$.

\vspace{3mm}
(6) Quantum Niederreiter PKE\\

$g(m,r)=m\oplus r$ and $f(m,r)=mH'^T$. The trapdoor $s\triangleq(M,H,P)$ satisfies $MHP=H'$.

\vspace{3mm}
(7) Quantum Okamoto-Tanaka-Uchiyama PKE\\

$g(m,r)=m\oplus r$ and $f(m,r)=\sum_{i=1}^ne_ib_i$, here $e_1\cdots e_n$ is the constant weight encoding of $m$. The trapdoor is $s\triangleq(g,d,p,p_1,p_2,\ldots,p_n)$.

\vspace{3mm}
In these seven QPKE protocols, the protocols (2) and (4) satisfy the case related with Formula.(\ref{equ0})(\ref{equ1}).
In these two protocols, a classical message is transferred and the value of $r$ is not computed
during the decryption process. We can see that the other protocols satisfy the case related with Formula.(\ref{equ2}).
No classical information is transferred in these protocols, and $r$ is computed during the decryption process.

\subsection{An authentication protocol~\cite{Yang03}}
Consider the original SN-S authentication scheme~\cite{Safavi91}.
Suppose generator matrix $G_s$ is a $k$ by $n_1$ matrix and in standard form:
$G_s=[I_k|A]$, here $I_k$ is the $k$ by $k$ identity matrix,
$A$ is chosen randomly from $k$ by $n_1-k$ matrices. The $[n_1,k]$ linear code generated by $G_s$ need not be of any error-correcting or error-detecting capability. Generalized inverse matrix $G_s^{-1}$ satisfies: $G_sG_s^{-1}=I_k$. Suppose the parity check matrix of the linear code generated by $G_s$ is $H_s$, then $H_s=\left[-A^T|I_{n-k}\right]$. Public-key authentication of quantum message is proposed in the following steps.

(1) Alice encodes a $k$-qubit message $\sum_m\alpha_m|m\rangle$ into $n_1$-qubit one as follows:
\begin{eqnarray}
&&\sum_m\alpha_m|m\rangle|0\rangle \nonumber\\
&\rightarrow& \sum_m\alpha_m|m\rangle|mG_s\rangle \nonumber\\
&\rightarrow& \sum_m\alpha_m|m\oplus mG_sG_s^{-1}\rangle|mG_s\rangle=|0\rangle\sum_m\alpha_m|mG_s\rangle.
\end{eqnarray}

(2) Alice uses Bob's public-key $G'$ to encrypt $n_1$-qubit state $\sum_m\alpha_m|mG_s\rangle$ via Quantum McEliece PKE.

(3) Bob uses his private-key $(S,G,P)$ to decrypt the received quantum state and obtains the $n_1$-qubit plaintext $\sum_m\alpha_m|mG_s\rangle$.

(4) Bob performs the following transformations on the quantum state $\sum_m\alpha_m|mG_s\rangle$.
\begin{eqnarray}
&&|0\rangle\sum_m\alpha_m|mG_s\rangle|0\rangle \nonumber\\
&\rightarrow& |0\rangle\sum_m\alpha_m|mG_s\rangle|mG_sG_s^{-1}\rangle=|0\rangle\sum_m\alpha_m|mG_s\rangle|m\rangle \nonumber\\
&\rightarrow& \sum_m\alpha_m|mG_sH_s\rangle|mG_s\rangle|m\rangle=|0\rangle\sum_m\alpha_m|mG_s\rangle|m\rangle \nonumber\\
&\rightarrow& |0\rangle\sum_m\alpha_m|mG_s\oplus mG_s\rangle|m\rangle=|0\rangle|0\rangle\sum_m\alpha_m|m\rangle.
\end{eqnarray}

(5) Bob measures the first register to check whether it is in the state $|0\rangle$. If it is, he accepts the message in the third register.

For the case that $G_s$ is public, the scheme is a public-key data integrity scheme.
This scheme can be modified to be one against substitution, the details are given in Sec.\ref{seccount3}.

\subsection{Quantum message signature protocols}
We have established a theoretical framework of signature of quantum message. Here, two protocols are proposed as the instances of the theoretical framework. One is not secure in post-quantum era, while the other is post-quantum secure.

In the first protocol, we take the function $f(x)=x^e\textrm{mod}N$ as the trapdoor one-way function, here the numbers $e$ and $N$ is the same as in Sec.\ref{seccount4}. Because $f(x)=x^e\textrm{mod}N$ is a trapdoor one-way permutation, it can be expressed as $f:\{0,1\}^k\times\{0,1\}^n\longrightarrow\{0,1\}^k\times\{0,1\}^n$,
here $k+n=\lceil \textrm{log}_2N\rceil$. That means, in the framework described in Sec.\ref{seccount6}, the
random number generated by Bob is $r_B\in\{0,1\}^k$ and the random number generated by Alice is $r_A\in\{0,1\}^n$.
Alice uses her private-key $d$ to compute $f^{-1}(r_B,r_A)=(r_B,r_A)^d\textrm{mod}N=(r,r')$, then obtains $r\in\{0,1\}^k$ and $r'\in\{0,1\}^n$. With the number $r$ and the function $f$, Alice signs the $n$-qubit message $\sum_m\alpha_m|m\rangle$ and gets $k+2n$-qubit state $\sum_m\alpha_m|m\rangle|(r,m)^e\textrm{mod}N\rangle$, then sends it to Bob. After receiving the quantum state, Bob tells Alice that he has received it. Then Alice announces $r$ and $r'$. Bob computes $(r,r')^e\textrm{mod}N$, and if its first $k$ bits are $r_B$, he performs the transformation
\begin{equation}
\sum_m\alpha_m|m\rangle|(r,m)^e\textrm{mod}N\rangle\longrightarrow\sum_m\alpha_m|m\rangle|0\rangle.
\end{equation}
Bob measures the second quantum register and accepts the signature if and only if the second register is in the state $|0\rangle$.

This signature protocol bases its security on the hardness of factoring problem. Because there exists efficient
quantum algorithm for this problem~\cite{Shor94}, the protocol is not secure in post-quantum era.

In the second protocol, we take the function $f(x)=x_1G'\oplus x_2$, here $x\in\{0,1\}^{k+n}$ is divided into two parts $x_1\in\{0,1\}^k$ and $x_2\in\{0,1\}^n$, and the $k\times n$ matrix $G'$ is the same as in Sec.\ref{seccount5}. Thus the trapdoor one-way function can be expressed as $f:\{0,1\}^k\times\{0,1\}^n\longrightarrow\{0,1\}^{\frac{n}{2}}\times\{0,1\}^{\frac{n}{2}}$. In the framework described in Sec.\ref{seccount6}, the random number generated by Bob is $r_B\in\{0,1\}^{\frac{n}{2}}$ and the random number generated by Alice is $r_A\in\{0,1\}^{\frac{n}{2}}$. It is required that $W_H(r_A)=W_H(r_B)=\lfloor\frac{t}{2}\rfloor$, here $W_H(x)$ denotes the Hamming weight of $x$, and $t$ is the correctable number of errors. Alice uses her private-key $s\triangleq(S, G, P)$ to compute $(r',r)$ which satisfy $r'G'\oplus r=(r_B,r_A)$, then obtains $r'\in\{0,1\}^k$ and $r\in\{0,1\}^n$. With the number $r$ and the function $f$, Alice signs the $k$-qubit message $\sum_m\alpha_m|m\rangle$ and gets $2k+n$-qubit state $\sum_m\alpha_m|m\rangle|mG'\oplus r\rangle$, then sends it to Bob. After receiving the quantum state, Bob tells Alice that he has received it. Then Alice announces $r$ and $r'$. Bob computes $r'G'\oplus r$, and if its first $\frac{n}{2}$ bits are $r_B$, he performs the transformation
\begin{equation}
\sum_m\alpha_m|m\rangle|mG'\oplus r\rangle\longrightarrow\sum_m\alpha_m|m\rangle|0\rangle,
\end{equation}
and measures the second quantum register. He accepts the signature if and only if the second register is in state $|0\rangle$.

For the second protocol, it is worth to mention that, in order to make it possible to compute $f^{-1}$ efficiently,
the sum of Hamming weights of $r_A$ and $r_B$ should not exceed $t$. Denote $H$ as the check matrix of the code
generated by $G$. If $r'G'\oplus r=(r_B,r_A)$, according to $(r'G'\oplus r)P^{-1}H=rP^{-1}H$, we have $rP^{-1}H=(r_B,r_A)P^{-1}H$. Because $P$ is a $n\times n$ permutation, $W_H(wP^{-1})=W_H(w)$ for any $w\in\{0,1\}^n$. Then $W_H(r)=W_H(r_B,r_A)=W_H(r_B)+W_H(r_A)$. Because $W_H(r)$ should not exceed $t$, the sum of Hamming weights of $r_A$ and $r_B$ should not exceed $t$ also. Here, we take $W_H(r_A)=W_H(r_B)=\lfloor\frac{t}{2}\rfloor$ for convenience.
\section{Security evaluation}
Now we evaluate the security of proposed theoretical frameworks.

{\bf Proposition 1:} In the QPKE framework based on induced trapdoor OWQT, it can be verified that the encryption transformation does not decrease the fidelity between two quantum states.

{\bf Proof:}
For two quantum messages $|M_1\rangle=\sum_m\alpha_m|m\rangle$ and $|M_2\rangle=\sum_m\alpha_m'|m\rangle$,
their fidelity is
\begin{equation}
F(|M_1\rangle,|M_2\rangle)=\left|\langle M_1|M_2\rangle\right|=\left|\sum_m\alpha_m^*\alpha_m'\right|.
\end{equation}
The ciphers of $|M_1\rangle$ and $|M_2\rangle$ are $\sum_rp_r\rho_r$ and $\sum_rp_r\sigma_r$ respectively, here $\rho_r$ and $\sigma_r$ can be expressed as
\begin{equation}\rho_r=(\sum_m\alpha_m|g(m,r)\rangle|f(m,r)\rangle)(\sum_m\alpha_m^*\langle g(m,r)|\langle f(m,r)|),\end{equation} and \begin{equation}\sigma_r=(\sum_m\alpha_m'|g(m,r)\rangle|f(m,r)\rangle)(\sum_m\alpha_m'^*\langle g(m,r)|\langle f(m,r)|).\end{equation}
According to the joint concavity of fidelity, it holds that
\begin{equation}
F\left(\sum_rp_r\rho_r,\sum_rp_r\sigma_r\right)\geq \sum_rp_rF\left(\rho_r,\sigma_r\right).
\end{equation}
Because $\rho_r$ and $\sigma_r$ are pure states, then
\begin{eqnarray}
F(\rho_r,\sigma_r)&=&\left|\sum_m\sum_n\alpha_m^*\alpha_n'\langle g(m,r)|g(n,r)\rangle\langle f(m,r)|f(n,r)\rangle\right| \nonumber\\
&=&\left|\sum_m\alpha_m^*\alpha_m'\right|=F(|M_1\rangle,|M_2\rangle).
\end{eqnarray}
Therefore, $F(\sum_rp_r\rho_r,\sum_rp_r\sigma_r)\geq F(|M_1\rangle,|M_2\rangle)$.~$\Box$

From this proposition, we can also know that the trace distance between two quantum states does not
increase after the encryption transformation. It can be seen that the holding of these results relates with the fact that
the encryption transformation can be regarded as a trace-preserving quantum operation to Bob and Eve.

%
According to the definition of induced trapdoor OWQT, the function $f(m,r)$ and $g(m,r)$ are classical functions.
Finding the trapdoor $s$ is a classical computational problem in each protocol. Thus, the QPKE protocols
based on induced trapdoor OWQT are just computational secure.

Now we prove that those seven encryption protocols are at least as secure as their classical counterparts.

{\bf Theorem 2:} The quantum McEliece PKE is more secure than classical McEliece PKE.

{\bf Proof:}
Suppose there is a quantum algorithm $A$, which can efficiently transform the
cipher state $\sum_m\alpha_m|mG'\oplus r\rangle$ into quantum message $\sum_m\alpha_m|m\rangle$.
In order to decrypt arbitrary classical cipher $m_0G'\oplus r_0$,
we firstly prepare a quantum state $|m_0G'\oplus r_0\rangle$.
Then, the quantum state $|m_0G'\oplus r_0\rangle$ is an input to the quantum algorithm $A$,
and will be transformed into the quantum state $|m_0\rangle$. Finally, the classical message $m_0$ is obtained via measuring the output quantum state $|m_0\rangle$. Thus, if there is an attack to quantum McEliece PKE, there would be an attack to classical McEliece PKE.

However, an attack to classical McEliece PKE does not mean an attack to quantum McEliece PKE.
There are several kinds of attack to classical McEliece PKE, such as Korzhik-Turkin attack~\cite{Korzhik91},
message-resend attack and related-message attack~\cite{Berson97}. Since the detail of Korzhik-Turkin attack has not
been given till now, the efficiency of this attack is still an open problem. Because iterative decoding algorithm
is used in the Korzhik-Turkin attack, and quantum state cannot be reused, it fails when attacking quantum McEliece PKE.
Though classical McEliece PKE have to be improved to prevent message-resend attack
and related-message attack~\cite{Sun98}, these attacks also fail while facing the quantum McEliece PKE.


Therefore, quantum McEliece PKE is more secure than classical McEliece PKE.~$\Box$

In the same way, it can be proved that the other QPKE protocols within our framework are at least as secure as their classical counterparts.

In our framework of authentication, QPKE scheme are used to ensure the quantum message with authentication being
transmitted securely. Eve cannot get the quantum message with authentication if she cannot break related QPKE scheme. So it seems hard for her to successfully break the integrity of quantum message.

In our framework of digital signature, if Eve wants to forge the signature of Alice,
she must capture the number $r_B$ and find $(r,r')$ which satisfies $f(r,r')=(r_B,*)$. However, this implies she can invert the trapdoor one-way function $f$. So the security of digital signature is ensured by the trapdoor one-way function $f$.

\section{Discussions}
(1) In the framework of QPKE, given the random number $r$ or the trapdoor information $s$ of $f(m,r)$,
the transformation from cipher state $\sum_m\alpha_m|g(m,r)\rangle|f(m,r)\rangle$ to plaintext
state $\sum_m\alpha_m|m\rangle$ can be completed efficiently. Both $r$ and $s$ are trapdoors of the induced trapdoor OWQT $U_{fg}(r)$.
Moreover, it can be concluded within the framework that, as an encryption algorithm is one with random number,
the disentanglement in the decryption is a process of extracting the pure state from the received mixed state.

(2) If the message to be encrypted is $\sum_m\alpha_m|m\rangle$, and only one of $\alpha_m$ is 1 and others are 0,
the QPKE protocols above degenerate into corresponding classical PKE protocols respectively.

(3) The encryption transformations in this paper are trace-preserving quantum operation to Bob and Eve, which are induced from the classical functions
$g(m,r)$ and $f(m,r)$. So that our protocols can be regarded as ones constructed via trace-preserving quantum operations.

(4) Our QPKE schemes are designed to encrypt quantum message $\sum_m\alpha_m|m\rangle$.
However, if we consider the number $r$ involved as classical message to be encrypted, the QPKE schemes can
also transmit classical information via sending quantum states, so this kind of QPKE scheme
can also be named as "quantum envelope". In addition, since the attacks to classical McEliece PKE,
such as Korzhik-Turkin attack~\cite{Korzhik91}, message-resend attack and related-message attack~\cite{Berson97},
fail to attack quantum McEliece PKE, we believe
it is more secure to transmit classical information via quantum McEliece PKE than via classical McEliece PKE.

(5) It can be seen that our QPKE schemes are computationally secure.
The protocols in Sec.\ref{seccount1} base their security on factoring problem or discrete
logarithms problem, so they are not secure in post-quantum era. However, since the protocols in Sec.\ref{seccount2}
base their security on the hardness of different NP-complete problems, we guess they are secure against quantum attacks.

\section{Conclusions}
Induced trapdoor OWQT has been introduced, and a theoretical framework of QPKE based on
it has been proposed. Seven QPKE protocols are given within this framework, such as quantum version of RSA, ElGamal, Goldwasser-Micali, elliptic curve, McEliece,
Niederreiter and Okamoto-Tanaka-Uchiyama PKE. These QPKE protocols for quantum message are shown to be at least as secure as
their classical counterparts. The last three protocols may be secure under the assumption that NP-complete problems
cannot be solved efficiently with quantum algorithms. Besides, theoretical frameworks for public-key authentication
and signature of quantum message are also proposed. A public-key authentication protocol and two digital signature protocols are given as their instances.

\section*{Acknowledgements}
This work was supported by the National Natural Science Foundation of
China under Grant No. 60573051.


\end{document}